\documentclass[%
 reprint,
 amsmath,amssymb,
 aps,
]{revtex4-1}

\usepackage{graphicx}
\usepackage{dcolumn}
\usepackage{bm}
\usepackage{hyperref}
\usepackage[dvipsnames]{xcolor}
\usepackage{bm}
\usepackage{bbold}
\usepackage{bbm}

\begin{document}

\title{Proposal for THz lasing from a topological quantum dot}

\author{Marie Rider}
\email{marie.rider@ic.ac.uk}
\affiliation{The Blackett Laboratory, Imperial College London, SW7 2AZ}

\author{Vincenzo Giannini}
\homepage{www.GianniniLab.com}
\affiliation{Instituto de Estructura de la Materia (IEM-CSIC), Consejo Superior de Investigaciones Cient{\'i}ficas, Serrano 121, 28006 Madrid, Spain}
\affiliation{Technology Innovation Institute, Building B04C, Abu Dhabi P.O. Box 9639, United Arab Emirates}
\date{\today}
\begin{abstract}
Topological quantum dots (TQDs) are 3D topological insulator nanoparticles with radius below 100 nm, which display symmetry-protected surface states with discretized energies. We propose a scheme which harnesses these energy levels in a closed lasing scheme, in which a single TQD, when pumped with incoherent THz light, lases from its surface states in the THz regime. The time scales associated with the system are unusually slow, and lasing occurs with a very low threshold. Due to the low threshold, we predict that blackbody radiation at room-temperature provides enough photons to pump the system, providing a route to room-temperature THz lasing with blackbody radiation as the pump. 
\end{abstract}
\maketitle

\section{Introduction}
\label{sec:intro}

The creation of robust, low threshold sources in the terahertz (THz) gap ($\sim$ 0.1-10 THz) is an important frontier in modern applied physics and this rapidly expanding field has recently received much attention. With much needed applications in bio-medicine~\cite{yang2016biomedical}, wireless communications~\cite{kleine2011review} and security technologies~\cite{federici2005thz}, THz waves penetrate materials opaque to other wavelengths, while posing only minimal risks due to their non-ionizing behaviour (unlike for example, x-rays). Academic applications range from molecular spectroscopy~\cite{baxter2011terahertz} to sub-millimetre astronomy~\cite{pilbratt2010herschel}, influencing physics on vast length scales.

Topological quantum dots (TQDs) are 3D topological insulator nanoparticles with radius $\leq 100$ nm, whose topological surface states undergo quantum confinement resulting in a discretised Dirac cone, with energy levels which increase in degeneracy away from the Dirac point. We avoid the acronym TIQD which is generally used for non-equiaxial TI quantum dots~\cite{cho2012topological,huang2019thz,wu2017spin,zhang2021interference,PhysRevLett.121.256804}. This work focuses on equiaxial 3D TI nanoparticles, which is a small but steadily expanding field~\cite{rider2019perspective,rider2020experimental,siroki2016single,gioia2019spherical,imura2012spherical,paudel2013three,castro2020optical,PhysRevB.101.165410}. Much like other types of quantum dots and atomic systems, the discrete energy levels of the TQD make it a natural system for controlled interactions with light. TQDs have energy levels separated by THz frequencies, making them a potential route to THz lasing. 

Success has been found in other THz systems, such as solid-state and molecular systems~\cite{gousev1999widely,del2011terahertz,kumar20111,pagies2016low,chestnov2017terahertz,zeng2020photonic,chevalier2019widely}, metasurfaces~\cite{keren2019generation}, quantum gases~\cite{chevalier2019widely} and more recently air plasma~\cite{koulouklidis2020observation}. The use of topological systems as a robust platform for lasing is also developing as an exciting field, most notably in the mid-infrared ~\cite{st2017lasing,bahari2017nonreciprocal, klembt2018exciton,malzard2018nonlinear,harari2018topological,bandres2018topological,parto2018edge,zhao2018topological,ota2018topological,kartashov2019two,ozawa2019topological}, although work has been done on THz lasers using 2D topological insulator quantum dots (TIQDs), requiring electron injection and an IR pump~\cite{huang2019thz}. 

In this work we use Monte Carlo (MC) simulations to demonstrate that for a model system of a single TQD in an open cavity at zero temperature, the TQD states can be pumped and act as an active gain medium, emitting coherent light in the THz. We show that lasing from a single TQD is possible with very low threshold. One of the major technological distinctions between the TQD laser and other THz lasers is that we do not pump between bulk bands (as done in molecular THz lasers and TIQD lasers), but instead we employ transitions within the Dirac cone to both pump and lase. This is possible due to the energy level configuration and selection rules of the TQD~\cite{paudel2013three,gioia2019spherical}. An important consequence of THz pumping is that the density of thermal, room temperature photons at this frequency (as opposed to those present at optical and IR frequencies) would be high enough that, in principle, an external pumping source beyond blackbody radiation may not be needed. This feature, alongside the unusual time scales in the system and the robustness of the TQD states make for a novel implementation. Experimental progress in producing TI nanostructures such as nanowires~\cite{peng2010aharonov,kong2010topological,tian2013dual,hong2014one,liu2020development} and finite thickness nanodisks and nanoplates~\cite{min2012quick,li2012controlled,veldhorst2012experimental,vargas2014changing,liu2015one,dubrovkin2017visible} has been very successful, and there has been recent progress with equiaxial 3D TI nanoparticles~\cite{rider2020experimental}. As well as adding another tool to the tool-box of THz lasing, TQDs could also find use in other technological applications, such as quantum memory and quantum computing~\cite{paudel2013three}, and as components of hybrid systems, such as coupled to semiconductor quantum dots~\cite{castro2020optical} and quantum emitters~\cite{PhysRevB.101.165410}.

This paper is structured as follows: In Section~\ref{sec:TQDs} we will briefly review the theory of spherical topological insulator nanoparticles and in Section~\ref{sec:optical} discuss their optical properties. In Section~\ref{sec:kp} we will describe how higher order corrections to the Dirac cone become important when a TI nanoparticle is placed in a high quality cavity, and how this can be utilised to create a closed scheme of energy levels. We will then describe in Section~\ref{sec:MC} how Monte Carlo simulations are used to capture the dynamics of the system when pumped and in Section~\ref{sec:lasing} give results demonstrating that a single TI nanoparticle can be used as a THz laser at zero temperature. We discuss how the low threshold of this system could allow for the utilisation of room temperature thermal photons as the pumping source, and conclude in Section~\ref{sec:outlook} with an outlook. 
%

\section{3D topological insulator quantum dots (TQDs)}
\label{sec:TQDs}
\begin{figure}[h!]
    \centering
   \includegraphics[width=\columnwidth]{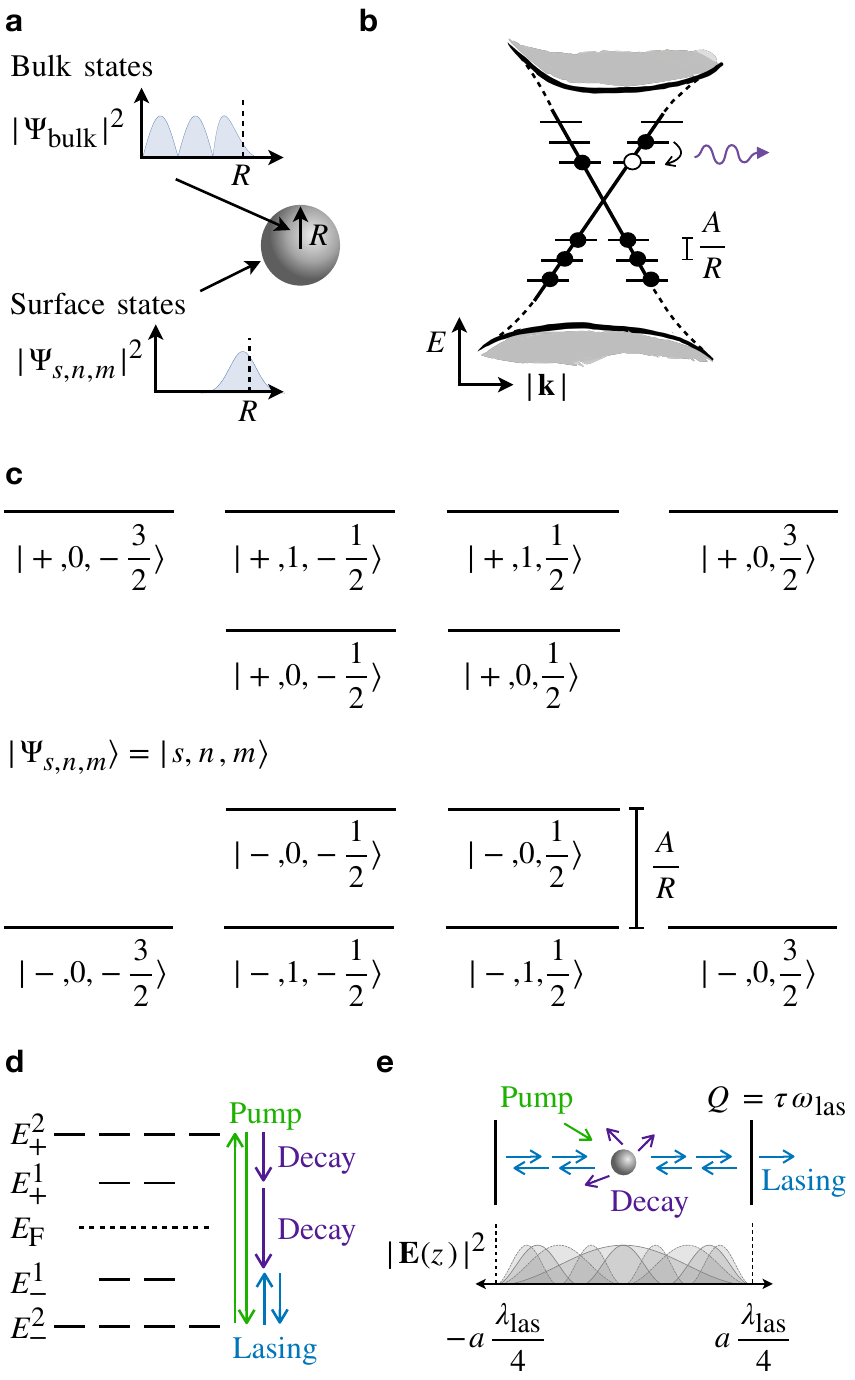}
    \caption{\textbf{Lasing topological quantum dots}: \textbf{(a)} Schematic of topological quantum dot, showing bulk and surface states. \textbf{(b)} Discretization of Dirac cone for small nanoparticles, with energy level separation near the Dirac point inversely proportional to particle radius, $R$. Transitions between energy levels are facilitated by the absorption or emission of THz light. \textbf{(c)} The discrete energy levels are labelled by three quantum numbers $|\mathrm{s,n,m}\rangle$. We present the labels for the four energy levels closest to the Dirac point. \textbf{(d)} Example lasing scheme with interband pumping and bottom-most transition as the lasing transition. Fermi level $E_{\mathrm{F}}=0$. \textbf{(e)} Schematic of TQD in cavity of length $a \lambda_\mathrm{las}/2$ with $a$ an integer, lasing frequency $\nu_{\mathrm{las}}$, and the $c$-axis of the material aligned with the cavity axis. Spatial dependence of cavity electric field gives spatial dependence of DOS and consequently transition rates. TQD placed at centre of cavity for maximal interaction with the cavity field, meaning that only modes with $a$ odd will have non-zero transition rates in the cavity axis.}
    \label{fig:schematic}
\end{figure}
Topological insulator (TI) nanoparticles of radius $5\leq R/\mathrm{nm} \leq 100$ support symmetry-protected surface states with quantised energy levels. For spherical nanoparticles~\cite{imura2012spherical, gioia2019spherical,paudel2013three,siroki2016single} (presented schematically in FIG.~\ref{fig:schematic}a), close to the $\Gamma$-point the wavefunctions for these surface states are given by
\begin{align}
\begin{split} \label{eq:wavefunction}
     \Psi_{s, n, m} (\xi,\varphi) & = \langle \xi,\varphi | \Psi_{s,n,m} \rangle \\
     &= \frac{e^{im\varphi}}{2\sqrt{\pi }R}  N_{n,m}\chi_{s,n,m} (\xi), 
\end{split}
\end{align}
where
\begin{align}
  \chi_{s,n,m} (\xi)    &= \begin{pmatrix} (1-\xi)^{\frac{1}{2}\sigma^-} (1+\xi)^{\frac{1}{2}\sigma^+}J_n^{\sigma^- \sigma^+}(\xi)\\-\frac{sm}{|m|} (1-\xi)^{\frac{1}{2}\sigma^+} (1+\xi)^{\frac{1}{2}\sigma^-}J_n^{\sigma^+ \sigma^-}(\xi)\end{pmatrix} ,
\end{align}
and $\sigma^{\pm} = |m\pm\frac{1}{2}|$, $\xi = \mathrm{cos}\vartheta$, $s = \pm$, $n = 0,1,2, ...$, $m = \pm 1/2,3/2,5/2,... $ and $J_{n}^{\sigma^{\pm}\sigma^{\mp}}$ are Jacobi polynomials~\cite{imura2012spherical}. $N_{n,m}$ is a normalisation factor (given in Supplementary Material~\ref{app:jacobi}). The corresponding discretized energy levels of these states are given by 
\begin{align}
    E_{s, n, m} & = \frac{s A}{R}\left(n + |m|+ \frac{1}{2} \right) ,
\end{align}
where $A$ is a material-dependent constant found from \textit{ab initio} calculations~\cite{liu2010model}. $A$ relates to the strength of spin-orbit coupling (SOC) in the material, averaged over the three Cartesian axes. This approximation of isotropic SOC allows for the analytical description of the topological surface states and their energy levels. Close to the $\Gamma$-point the energy levels are linearly spaced, and increase in degeneracy away from the Dirac point (i.e. $2,4,6,8,... $) as illustrated in FIG.~\ref{fig:schematic}b. The energy values are inversely proportional to $R$, and for large $R$ we recover the expected continuum Dirac cone. Examples of the energy levels associated with quantum numbers $(s,n,m)$ are given in FIG.~\ref{fig:schematic}c. The system of discrete energy levels can be tuned with particle size (much like the energy levels of a semiconductor quantum dot or topological insulator quantum dot), encouraging us to refer to this system as a topological quantum dot (TQD). The key differences between TQDs and semiconductor quantum dots are (i) the delocalization of states for TQDs vs the relatively localized states of the semiconductor counterpart, (ii) the quantisation of surface states rather than bulk states, which in turn affects (iii) the frequency-range of operation, which for semiconductor quantum dots is usually in the visible range and near infrared, but for TQDs is in the THz, e.g. for a Bi$_2$Se$_3$ TQD of $5<R/\mathrm{nm}< 100$, $A=3.0$ eV\AA~\cite{liu2010model} and the spacing between two energy levels is $0.72\mbox{-}14.4$ THz, contrasting with a band gap of $220$ meV $\sim 53.2$ THz. 

The high tunability of this system, combined with the unusual degeneracy and spacing of the energy levels suggests an interesting nanophotonic system~\cite{siroki2016single,rider2019perspective,rider2020experimental}. The goal of this paper will be to demonstrate that by pumping the electrons of the surface states between discrete energy levels while placed in a cavity (described schematically in Figures~\ref{fig:schematic}d and e), we can use a TQD as a THz laser. With this in mind, we will now discuss the TQD optical properties.  
%

\section{Optical properties of TQDs}
\label{sec:optical}
\begin{figure}
    \centering
    \includegraphics[width=\columnwidth]{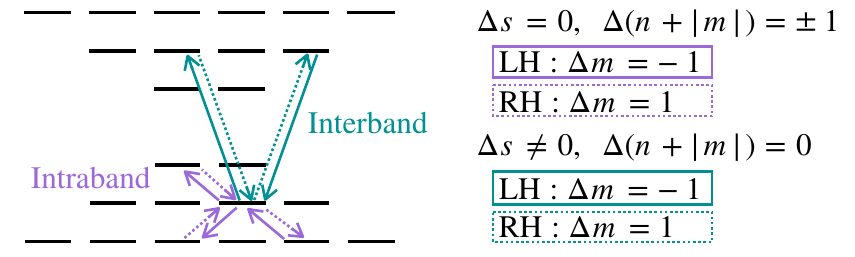}
    \caption{\textbf{TQD selection rules}: E1 selection rules for TQD state $|-,1,1/2\rangle$. Interband (green) transitions couple energy levels above and below the Dirac point. Intraband (purple) transitions couple levels within the same section of the Dirac cone.}
    \label{fig:selection}
\end{figure}

For a TQD with material $c$-axis aligned along the $z$-axis, the electric dipole (E1) selection rules for allowed optical transitions between TQD energy levels such that $|\Psi_{\mathrm{i}}\rangle \rightarrow |\Psi_{\mathrm{f}}\rangle$, where $\mathrm{i}=(s,n,m)$ and $\mathrm{f}=(s',n',m')$ can be found by examining the E1 matrix element (with some explicit examples given in Supplementary Material~\ref{app:transitions}),
\begin{align}
    V_{\mathrm{i,f}}(\omega_{\mathrm{i,f}}) = \langle \Psi_{\mathrm{f}}  | \mathbf{r} \cdot \bm{\epsilon}| \Psi_\mathrm{i} \rangle, 
    \label{eq:matrix_element}
\end{align}
where $\bm{\epsilon}$ is the polarisation vector of the incoming light. For stimulated processes, the selection rules for incoming right-hand (RH) polarised light along the $z$-axis are given by 
\begin{align}
    &\Delta s = 0,\quad\Delta (n+|m|) = \pm 1,\quad \Delta m = 1,  \\
    &\Delta s \neq 0,\quad\Delta (n+|m|) = 0, \quad \Delta m = 1.
\end{align} 
For incoming left-hand (LH) polarised light along the $z$-axis, the selection rules are
\begin{align}
    &\Delta s = 0,\quad\Delta (n+|m|) = \pm 1,\quad \Delta m = -1,  \\
    &\Delta s \neq 0,\quad\Delta (n+|m|) = 0, \quad \Delta m = -1.
\end{align} 
FIG.~\ref{fig:selection} illustrates the allowed stimulated transitions for a surface state with energy $E_{s,n,m}=-2A/R$ when excited with either LH or RH polarized light. Note that there are two distinct types of transitions: (i) interband transitions which couple the upper and lower halves of the Dirac cone requiring energy $\Delta E = 2E_{s,n,m}$ (green arrows), and (ii) intraband transitions which couple nearest energy levels, requiring $\Delta E = A/R$ (purple arrows). The transition rates of allowed stimulated E1 transitions are given by 
\begin{align}
    \Gamma_{\mathrm{i\rightarrow f}}^{\mathrm{stim}} = 2 \pi^2 \alpha \frac{c}{\hbar}  \bar{u}(\omega_{\mathrm{i,f}}) |V_{\mathrm{i,f}} |^2,
    \label{eq:stim_rate}
\end{align}
where $\bar{u}(\omega_{\mathrm{i,f}})=\hbar \omega_{\mathrm{i,f}} \bar{n}(\omega_{\mathrm{i,f}})  \rho(\omega_{\mathrm{i,f}})$, is the energy density at frequency $\omega_{\mathrm{i,f}}=2\pi \nu_{\mathrm{i,f}}$ and $\bar{n}(\omega_{\mathrm{i,f}})$ is the average number of photons in mode $\omega_{\mathrm{i,f}}$. We have assumed symmetry in photon polarisation such that $n_{\mathrm{LH}}(\omega_{\mathrm{i,f}})=n_{\mathrm{RH}}(\omega_{\mathrm{i,f}}) = \bar{n}(\omega_{\mathrm{i,f}})/2$. $V_{\mathrm{i,f}}$ is the matrix element given in Equation~\ref{eq:matrix_element}. The density of states in free space is $\rho_{\mathrm{free}}(\omega_{\mathrm{i,f}})= \omega_{\mathrm{i,f}}^2/\pi^2 c^3$, and in the cavity-axis of an open cavity of quality factor $Q$, formed of two parallel mirrors separated by a distance $L$,
\begin{align}
    \rho_{\mathrm{cav}}(\omega_{\mathrm{i,f}})= \frac{2}{\pi L} \frac{Q \omega_{\mathrm{cav}}}{\omega_{\mathrm{cav}}^2+4 Q^2 (\omega_{\mathrm{i,f}}-\omega_{\mathrm{cav}})^2}. 
\end{align}
The rate of spontaneous emission is given in free space by 
\begin{align}
    \Gamma^{\mathrm{spon}}_{\mathrm{i}\rightarrow \mathrm{f}}  =  \frac{4 \alpha \omega_{\mathrm{f,i}}^3 }{ 3 c^2}|\langle \Psi_{\mathrm{f}} | \mathbf{r} | \Psi_{\mathrm{i}} \rangle |^2,
    \label{eq:spon_rate_free}
\end{align}
and in the open cavity by 
\begin{align}
    \Gamma^{\mathrm{spon}}_{\mathrm{i}\rightarrow \mathrm{f}}  &= \frac{8 \pi \alpha c}{L} \frac{Q \omega_{\mathrm{i,f}}\omega_{\mathrm{cav}}}{\omega_{\mathrm{cav}}^2+4 Q^2 (\omega_{\mathrm{i,f}}-\omega_{\mathrm{cav}})^2} |\langle \Psi_{\mathrm{f}} | \mathbf{r}| \Psi_{\mathrm{i}}\rangle|^2.
    \label{eq:spon_rate_cavity}
\end{align}
Equations~\ref{eq:spon_rate_free} and \ref{eq:spon_rate_cavity} are polarisation independent. For a Bi$_2$Se$_3$ nanoparticle of radius $R=50$ nm in the linear approximation, the smallest frequency coupling states within the quantised Dirac cone is $\nu = 1.44$ THz and the spontaneous emission rate in free space for the intraband transition between levels $|+,1,-1/2 \rangle$ and $|+,0,1/2 \rangle$ is $ 7.0\cdot 10^3$ s$^{-1}$. Interband transitions are typically much faster, with the spontaneous transition in free space from $|+,1,-1/2 \rangle$ to $|-,1,1/2 \rangle$ given by $1.4 \cdot 10^5$ s$^{-1}$, with transition frequency $5.76$ THz. These rates are much slower than comparable processes in semiconductor dots, whose spontaneous emission rates are typically in the range $10^6$--$10^9$ s$^{-1}$ depending on their structure~\cite{bera2010quantum,jacak2013quantum}. This is due to the frequency dependence of the transition rates, with typical energy level separation in semiconductor dots much greater than in TQDs. 
%

%

\section{Higher order corrections}
\label{sec:kp}
\begin{figure}
    \centering
    \includegraphics[width=\columnwidth]{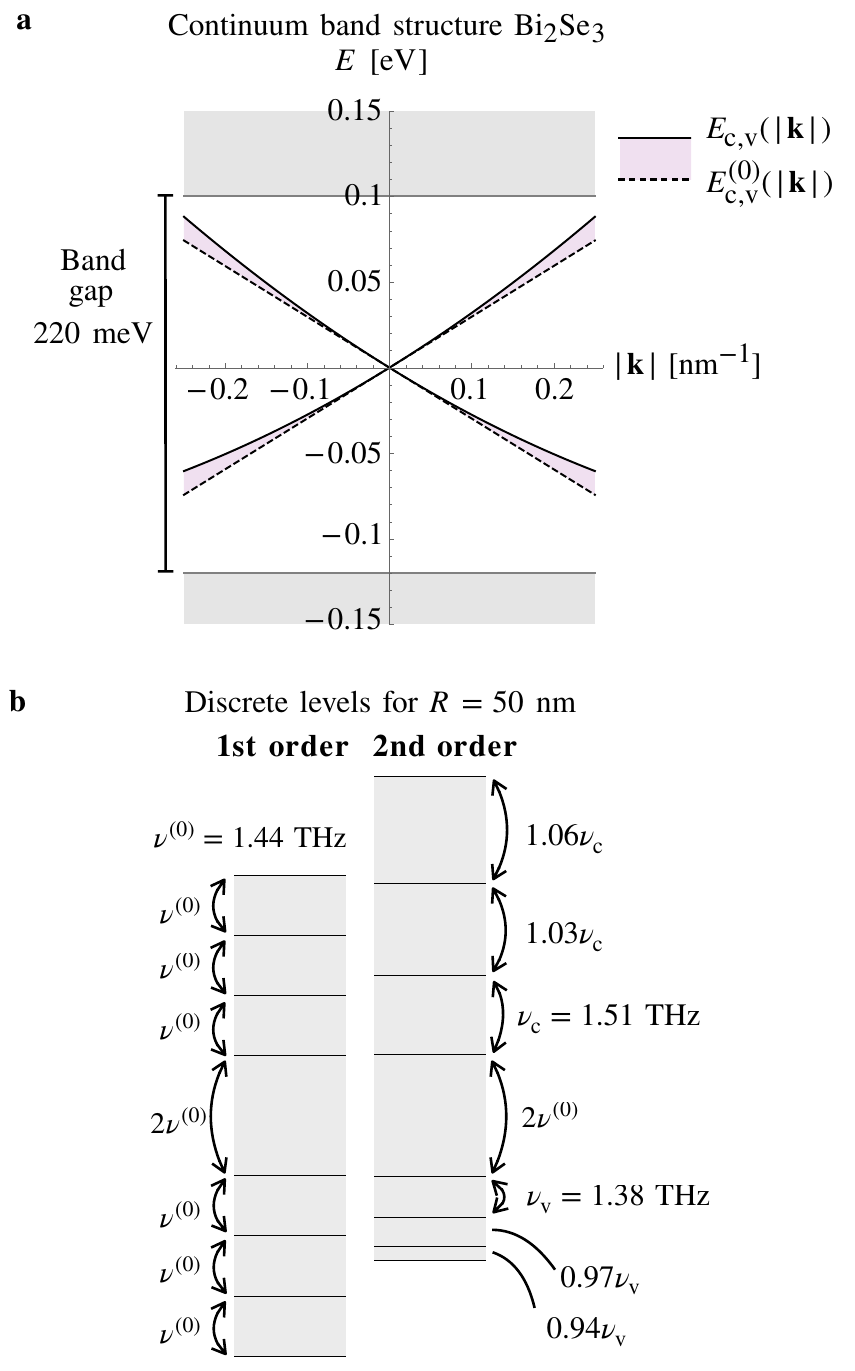}
    \caption{\textbf{k.p breakdown}: \textbf{(a)} Continuum band structure for Bi$_2$Se$_3$ with Dirac cone seen in the band gap (=220 meV~\cite{martinez2017determination}). Band structure up to and including $k^2$ terms (sold black line) deviates from the linear $k\cdot p$ approximation (dotted black line) away from the Gamma point.  \textbf{(b)} For a $R=50$ nm nanoparticle, discretised energy levels are expected. In the linear approximation, transition frequency $\nu^{(0)} = 1.44$ THz, whereas when taking into account the $k^2$ term, sequential transitions away from the $\Gamma$-point deviate increasingly.}
    \label{fig:kp_breakdown}
\end{figure}

Close to the $\Gamma$-point the surface state dispersion relation is described well by linear $\mathbf{k}\cdot \mathbf{p}$ theory and taken to be linear, and so the energy levels of the TQD surface states can be considered equally spaced. When higher order terms are to be taken into account, we return to the continuum model and use the dispersion relation
\begin{align}
    E_{\mathrm{c},\mathrm{v}} = \pm A |\mathbf{k}| + A_1 |\mathbf{k}|^2,
\end{align}
where $E_\mathrm{c,v}$ give the energies of the conduction and valence bands respectively. For Bi$_2$Se$_3$, $A=3.0$~eV\AA, $A_1 = 22.2$~eV\AA$^2$, and for Bi$_2$Te$_3$, $A=2.0$~eV\AA, $A_1 = 35.3$~eV\AA$^2$~\cite{liu2010model}. More details on these values are given in Supplementary Material~\ref{app:correction_k2}. On confinement of the surface states, $|\mathbf{k}| \sim 1/R$ and so the correction is proportional to $1/R^2$, demonstrated in FIG.~\ref{fig:kp_breakdown}a. For a Bi$_2$Se$_3$ TQD of $R=50$ nm, without the correction all nearest-level transitions occur with frequency $\nu^{(0)} = 1.44$ THz (asides from the transition coupling levels directly above and below the $\Gamma$-point, which couples at frequency $2\nu^{(0)}$), as illustrated in FIG.~\ref{fig:kp_breakdown}b. When taking the correction into account, transition frequencies in both the valence and conduction bands rapidly become off-resonant with the first transition in each band. The first and second transition frequencies in the upper Dirac cone show a disparity of 3$\%$. The first and third transition frequencies differ by 6$\%$, as shown in FIG.~\ref{fig:kp_breakdown}b.  The first transition frequencies in the valence and conduction bands differ by $\sim 9\%$. Variation in frequency is more rapid in smaller particles, and in materials for which $A_1$ is large. This disparity can thus be increased by considering a TQD of smaller radius, or by using a different material such as Bi$_2$Te$_3$.  It should be noted that interband transitions are unaffected due to the cancellation of the second-order correction.

When placed in a cavity, higher-order corrections to the TQD Hamiltonian must clearly be considered as transitions between energy levels will rapidly become off-resonant with the cavity frequency if tuned to a single transition frequency. For a cavity with lifetime greater than the timescale of free space spontaneous emission due to intraband transitions, $Q=\omega_{\mathrm{las}} > \tau_{\mathrm{spon}} \omega_{\mathrm{las}}$, $\tau_{\mathrm{spon}} \approx 10^{-4}$ s for a $R = 50$ nm particle, so $Q > 10^8$. This exceedingly high $Q$ factor is due to the unusual combination of a high transition frequency and long timescale. It is clear that for a cavity of such high $Q$ factor, the line-width is much smaller than the energy level separation, and the only mode coupled to the cavity will be the single resonant transition, such that $\rho_{\mathrm{cav}}(\omega) = 2Q/\pi \omega_{\mathrm{i,f}}L$. All other transitions within the Dirac cone and in the direction of the cavity axis will be drastically suppressed and considered negligible. However, for an open cavity these transitions may still occur in all directions not directly parallel to the cavity axis. 

\section{Surface state dynamics using Monte Carlo (MC)}
\label{sec:MC}
We now explain how a Monte Carlo method is used to study the time-evolving surface state dynamics of a TQD interacting with the electromagnetic field in a cavity. 

We move to a Fock basis, in which the quantum state of the TQD can be described by 
\begin{align}
    | \Psi \rangle = \prod_{\alpha=1}^{2M} \otimes | N_\alpha \rangle ,
\end{align}
where we take a product state of the $M$ states above and below the Dirac point, with $2M$ states in total forming a closed system of energy levels. Each state $\alpha=(s,n,m)$ obeys fermionic occupation rules such that $N_\alpha$ = 0~or~1. The photon density for the pump $\bar{n}_{\mathrm{pump}}$ is kept constant, and the photon density for modes not resonant with the cavity are zero. Stimulated emission will only be triggered by photons in the cavity axis and thus coherent with the lasing mode, $\nu_{\mathrm{las}}$, which is defined by the first spontaneous emission of frequency $\nu_{\mathrm{las}}$. The cavity (with quality factor $Q$) enforces feedback of coherent $\nu_\mathrm{las}$  photons. 

For a given time step $dt$, the transition rates given in Equations \ref{eq:stim_rate}, \ref{eq:spon_rate_free} and \ref{eq:spon_rate_cavity} are converted to probabilities $p_{\mathrm{i},\mathrm{f}} = \Gamma_{\mathrm{i\rightarrow f}} dt \ll 1$, where for stimulated processes $\Gamma_{\mathrm{i \rightarrow f}}$ will dynamically change as $\bar{n}_{\mathrm{i,f}}$ evolves via the dynamics. The TQD couples to the cavity mode with a rate proportional to the spatially-dependent probability density of the standing wave cavity mode shown in FIG.~\ref{fig:schematic}e. For simplicity we consider a TQD at the centre of the cavity.

The initial conditions are set by occupying states up to the desired Fermi level $E_\mathrm{F}$. For a given time step, all electrons may transition between states with the probabilities calculated from the relations given.  At the end of the time step, it is checked that the transitions have obeyed fermionic occupation rules (such that there is maximum one electron in any given state). If the transitions have resulted in multiple occupancy of any state, the time step is reset and the system required to undertake the step again. When a physically allowed time step has been undertaken, the entire system updates and progresses to the next time step. In this way, the system evolves whilst observing fermionic occupation rules. A single evolution of the system gives a single quantum trajectory, while averaging the simulation over many iterations gives the average expected results of a single TQD, or if multiplied by $N$, gives the result of $N$ uncorrelated TQDs~\cite{kalos2009monte}.

\section{THz Lasing from a single TQD}
\label{sec:lasing}
%
\begin{figure*}
    \centering
    \includegraphics[width=\textwidth]{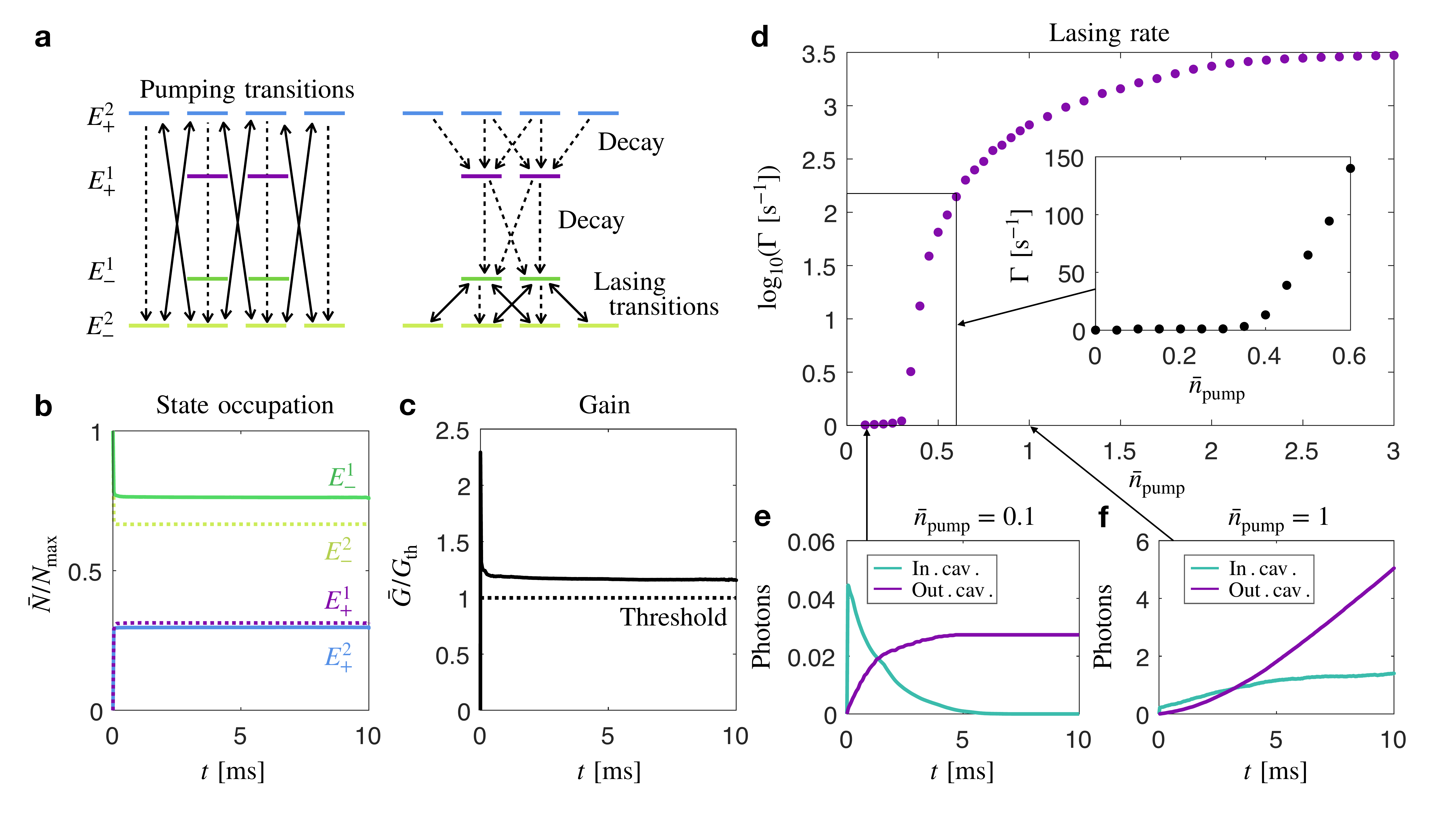}
    \caption{\textbf{TQD lasing}: \textbf{(a)} Schematic of all lasing scheme transitions. Solid lines denote stimulated processes (and downwards spontaneous processes) and dashed lines denote spontaneous processes only. \textbf{(b)} Evolution of normalised state occupation for $E_\mathrm{F}=0$, $\bar{n}_{\mathrm{pump}}=1$. Population inversion between energy levels $E_-^1$ and $E_-^2$ is demonstrated and \textbf{(c)} gain above threshold for this transition is demonstrated. \textbf{(d)} Steady-state lasing rate (coherent photons emitted from cavity per second) for varying $\bar{n}_{\mathrm{pump}}$ with inset showing low pumping regime $0\leq \bar{n}_\mathrm{pump}\leq 0.6$, with a lasing threshold of $\sim 0.35$. \textbf{(e)} shows number of coherent photons inside the cavity and emitted from the cavity as a function of $t$ for $\bar{n}_{\mathrm{pump}}=0.1$, which is below the lasing threshold and  \textbf{(f)} $\bar{n}_{\mathrm{pump}}=1$ which is above the lasing threshold.}
    \label{fig:lasing}
\end{figure*}
We now describe a specific closed lasing scheme using discrete surface states of the TQD at T=0. We start with a single Bi$_2$Te$_3$ TQD in a Fabry-Perot cavity with the material $c$-axis aligned with the cavity axis and Fermi level at $E_\mathrm{F}=0$. The TQD is pumped using an interband transition, and the cavity is tuned to the lowest intraband transition in the scheme, as illustrated in FIG.~\ref{fig:schematic}d. The arguments in section~\ref{sec:kp} show that all other transitions will be suppressed in the cavity axis. 
\\

We pump the system with photons of energy $E_+^2-E_-^2$, away from the cavity axis as shown in FIG.~\ref{fig:schematic}d and tune the cavity to the transition $\nu_\mathrm{las} = (E_-^1-E_-^2)/h$ by using an open Fabry-Perot cavity of length $L = a\lambda_\mathrm{las}/2$. For a $R = 50$ nm particle, this requires pumping with photons of frequency 5.76 THz and the lasing frequency will be 1.38 THz. We choose a cavity length $L=100.5\lambda_{las}\approx1.1$ cm. All processes not directly in the cavity axis will occur with the rate given in Equation~\ref{eq:spon_rate_free}, while in the cavity the stimulated rates will be given by Equation~\ref{eq:stim_rate} with density of states
\begin{align}
    \rho(\omega_{\mathrm{i,j}}) =\frac{Q}{2 \pi L \omega_{\mathrm{i,f}}} \delta(\omega_{\mathrm{i,j}}-2 \pi \nu_\mathrm{las}) ,
\end{align}
and the spontaneous rates in the cavity by  Equation~\ref{eq:spon_rate_cavity}. We place the TQD in the centre of the cavity at the peak of electric field density as displayed in FIG.~\ref{fig:schematic}e. All transitions in frequencies other than $\nu_{\mathrm{las}}$ are suppressed in the cavity axis. Only four energy levels are considered in the simulation as all other levels are decoupled, and all processes involved are depicted in FIG.~\ref{fig:lasing}a, with bold arrows indicating stimulated processes and dotted arrows indications spontaneous processes. For fixed $Q \sim 10^9$, the dynamics of the system are tuned only with the pumping rate, which in turn is controlled by $\bar{n}_{\mathrm{pump}} = \bar{n}\left(\omega_{\mathrm{pump}}\right)$ of the pumping light. For $\bar{n}_{\mathrm{pump}}$ = 1, the evolution of the normalised occupation of energy levels, $\bar{N}/N_{\max}$, is given in FIG.~\ref{fig:lasing}b. Steady state is quickly achieved, with clear population inversion between levels $E_-^2$ and $E_-^1$. Gain for the lasing transition is given by  
\begin{align}
   \bar{G} = \sum_{\mathrm{i<j}} \sigma_{\mathrm{j} \rightarrow \mathrm{i}} \left(\bar{N}_\mathrm{j} - \bar{N}_\mathrm{i} \right)\delta(\omega_{\mathrm{i,j}}-2 \pi\nu_{\mathrm{las}}),
\end{align}
where $\sigma_{\mathrm{j} \rightarrow \mathrm{i}} = \Gamma_{\mathrm{j} \rightarrow \mathrm{i}}/c \bar{n}_{\mathrm{i,j}}$ is the transition cross-section and $\bar{N}_\mathrm{i}$ is the average occupation of surface state $\mathrm{i}$.  $\bar{G}$ for $\bar{n}_{\mathrm{pump}}=1$ is plotted in  FIG.~\ref{fig:lasing}c. A maximum is achieved nearly instantaneously as electrons are excited from states with energy $E_-^2$ to $E_+^2$, creating population inversion in the lasing transition. Steady state is quickly achieved, with gain above the threshold value. Threshold gain is given by $G_{\mathrm{th}} = 1/\tau_\mathrm{las}c$, as internal losses are taken to be negligible in this simulation.

Displayed in FIG.~\ref{fig:lasing}d is the dependence of the lasing rate with varying $\bar{n}_{\mathrm{pump}}$. Lasing is found to occur for  $\bar{n}_{\mathrm{pump}} \geq 0.35$ (see inset of FIG.~\ref{fig:lasing}d). Below threshold, such as for $\bar{n}_{\mathrm{pump}}=0.1$ as demonstrated in FIG.~\ref{fig:lasing}e, coherent photons in the cavity (teal line) do not build up rapidly enough to compensate for photons being absorbed by the TQD or emitted from the cavity and so the number of cavity photons decays to 0. Lasing does not occur. Above threshold, such as for $\bar{n}_{\mathrm{pump}}=1$ displayed in FIG.~\ref{fig:lasing}f, a critical number of coherent photons build up in the cavity such that lasing can occur. The number of coherent photons in the cavity (teal line) slowly increases from 0 until a steady state is achieved. Coherent photons are emitted from the cavity (purple line), at a constant rate. The slope of this line gives the photon emission rate, $\Gamma \sim 6.6 \cdot 10^2~\mathrm{s}^{-1}$. The power of this lasing can be calculated by assuming photons are emitted over an area commensurate with the cross-section of the nanoparticle, such that $P = \Gamma h \nu_{\mathrm{las}}/ \pi R^2 \sim 7.7 \cdot 10^{-5}~\mathrm{W}/\mathrm{m}^2$. The incoming power is approximately $1.0~$W$/$m$^2$. The power conversion (power in vs power out) is roughly 0.0075\% efficiency. This is very low, but it should be remembered that we are considering a single TQD. The power output and thus efficiency should increase dramatically with increased number of TQDs.

Above the threshold value of $\bar{n}_{\mathrm{pump}}$ there is a small range in which the lasing rate exponentially increases, and then the lasing rate increases approximately linearly with increased $\bar{n}_{\mathrm{pump}}$. Competition between processes of different time-scales in the system (cavity emission rate, spontaneous emission rate and stimulated emission rate at the cavity frequency) eventually results in a slowing of the increase in lasing rate, and a maximum lasing rate of $\sim10^{3.5}$ s$^{-1}$. The lasing rates of this system have a very low threshold, but also occur at very slow times scales. This is due to the interesting interplay of length and energy scales in this system. While the lasing rates can be increased by the addition of more TQDs in the system, the system of a single TQD could offer interesting technological applications where the slow production of coherent photons is coveted. 

The simulations at $T=0$ described in this section were conducted for a $R = 50$ nm Bi$_2$Se$_3$ TQD requiring a pump with frequency 5.76 THz. At room temperature, the number of thermal photons in the pumping mode is $\bar{n}_{\mathrm{pump}} =0.66$, which is above the lasing threshold found for the proof-of-principle case. For $35\leq R/\mathrm{nm}\leq 100$, lasing frequency for the described setup is 0.70-1.93 THz, with a required pumping frequency of 2.88-8.23 THz. At room temperature, the number of photons in the pumping mode is $0.37 \leq n_{\mathrm{pump}}\leq 1.72$, which is above threshold for all cases. For the same lasing scheme but using Bi$_2$Te$_3$, with $35\leq R/\mathrm{nm} \leq 100$, the lasing frequencies achieved are $0.46-1.2$ THz, the pumping frequencies required are $1.94-5.54$ THz, and the number of photons per mode at room temperature at the pumping frequency is in the range $0.70-2.76$, which again is well above lasing threshold. 

All results presented so far have been for a single TQD. For multiple TQDs interacting via a single cavity mode (i.e. multiple TQDs aligned along the cavity axis), coherent photons emitted from one TQD will be available to trigger stimulated emission events in other TQDs. It is expected that the lasing rate will be amplified exponentially with increasing number of TQDs.
%

\section{Outlook}
\label{sec:outlook}

By considering higher-order corrections to the surface state Hamiltonian of a TQD, we have shown that a closed lasing scheme can be created such that a single TQD will produce coherent THz frequency light. Lasing occurs at a very low threshold, but also with an unusually slow time scale. We have demonstrated that population inversion can be achieved and that lasing can occur with a very low threshold - so low in fact, that the number of photons present in blackbody radiation at room-temperature is in theory enough to provide the pumping mechanism with no additional, external pumping source needed. This is only possible due to employing a THz pump. This concept deserves future investigation due to the important technological repercussions. To realistically demonstrate that lasing will occur at room temperature, we expect a 3D, closed cavity with a frequency-modulated Q factor will be required to control photon density and avoid thermal equilibrium. There is much work yet to be done to explore this system, but this scheme paves the way towards a novel room-temperature THz lasing source, with no external pump additional to blackbody radiation. 

\bibliographystyle{unsrt}
\bibliography{references}

\appendix

\section{Supplementary material}

\subsection{Jacobi polynomials}\label{app:jacobi}
Class of orthogonal polynomials  $J_n^{\alpha,\beta}(x)$, orthogonal w.r.t the weight $\left(1-x\right)^{\alpha}\left(1+x\right)^{\beta}$ on the interval $x \in [-1,1]$, where $\alpha, \beta >-1$, such that 
\begin{align}
\int_{-1}^1 \left( 1-x\right)^{\alpha}\left( 1+x\right)^{\beta} J_n^{\alpha\beta}(x)J_{n'}^{\alpha\beta}(x) dx = \frac{\delta_{nn'}}{N_{n\alpha \beta}^2} 
\end{align}
where 
\begin{align}
N_{n \alpha \beta}^2 = \frac{(2n+\alpha + \beta +1)!(n+\alpha + \beta)!n!}{2^{\alpha + \beta + 1} \Gamma (n+\alpha)!(n+\beta)!}.
\end{align}
\subsection{Matrix element}\label{app:transitions}
Some explicit values of the E1 matrix element $V_{\mathrm{i,f}}/R^2$ for LH polarized light (such that $\epsilon = (1,-i,0)/\sqrt{2}$) are given below.\\
\begin{centering}
\begin{tabular}{c c }
   $| \mathrm{i} \rangle \rightarrow | \mathrm{f} \rangle$ & $V_{\mathrm{i,f}}/R^2$ \\
   \hline
    $| +,0,-1/2 \rangle \rightarrow |+,1,1/2 \rangle $ & 1/9\\
    $| +,0,-1/2 \rangle \rightarrow |-,0,1/2 \rangle $ & 2/9\\
    $| +,0,-3/2 \rangle \rightarrow |+,0,-1/2 \rangle $ & 1/3\\
    $| +,0,-3/2 \rangle \rightarrow |-,1,-1/2 \rangle $ & 2/75
\end{tabular}
\end{centering}
 
\subsection{Surface state dispersion relation including higher order correction}
\label{app:correction_k2}
The four-band bulk Hamiltonian for a topological insulator, given in Reference~\cite{liu2010model} is given by 
\begin{align}
\begin{split}
\mathbf{H}(\mathbf{k})     
    = &~\epsilon_0 (\mathbf{k})\mathbbm{1}_2\otimes\mathbbm{1}_2 + M(\mathbf{k}) \mathbbm{1}_2\otimes \boldsymbol{\sigma}_{\boldsymbol{3}}   \\
    &+ A_0 k_x \boldsymbol{\sigma}_{\boldsymbol{1}} \otimes \boldsymbol{\sigma}_{\boldsymbol{1}} + A_0 k_y \boldsymbol{\sigma}_{\boldsymbol{2}} \otimes \boldsymbol{\sigma}_{\boldsymbol{1}}\\
     &+ B_0 k_z \boldsymbol{\sigma}_{\boldsymbol{3}} \otimes \boldsymbol{\sigma}_{\boldsymbol{1}},
\end{split}
\end{align}
where $M(\mathbf{k}) = M_0+M_1 k_z^2 + M_2 (k_x^2+k_y^2)$ and $\epsilon_0 (\mathbf{k}) = C_0 + C_1 k_z^2 +C_2 (k_x^2+k_y^2)$. For Bi$_2$Se$_3$, $A_0=3.33$~eV\AA, $B_0=2.26$~eV\AA, $C_0=-0.0083$~eV, $C_1=5.74$~eV\AA$^2$, $C_2=30.4$~eV\AA$^2$, $M_0=-0.28$~eV\AA$^2$, $M_1=6.86$~eV\AA$^2$ and $M_2=44.5$~eV\AA$^2$. The linear terms may be simplified by approximating spin-orbit coupling to be isotropic, such that $A=(A_0+2B_0)/3$ as used in the main text. The term $\epsilon_0(\mathbf{k})$ is often neglected to make the solution for the surface more tractable, however if this term is reintroduced then the leading $k^2$ terms in the surface state energies are given by $C_1 k_z^2+C_2(k_x^2+k_y^2)$. Averaging over the three Cartesian coordinates, for Bi$_2$Se$_3$ we arrive at the value $A_1=(C_1+2C_2)/3=22.18$~eV\AA$^2$ used in this work. 

\end{document}